\documentclass[preprint, aip]{revtex4-1}

\newcommand{\E}{\mathcal{E}}
\newcommand{\ed}[1]{\mathrm{d}{#1}}

\usepackage{mathtools}
\usepackage{upgreek}
\usepackage{amsmath}
\usepackage{amsfonts}

\begin{document}

\title{Geometric thermodynamics: black holes and the meaning of the scalar curvature}
\author{M. \'A. Garc\'ia-Ariza}
\email{magarciaariza@alumnos.fcfm.buap.mx}
\affiliation{Facultad de Ciencias F\'isico Matem\'aticas, Universidad Aut\'onoma de Puebla, Apartado Postal 1152, 72000, Puebla, Puebla, M\'exico}
\author{Merced Montesinos}
\email{merced@fis.cinvestav.mx}
\affiliation{Departamento de F\'{\i}sica, Cinvestav, Instituto Polit\'ecnico Nacional 2508, San Pedro Zacatenco, 07360, Gustavo A. Madero, Ciudad de M\'exico, M\'exico}
\author{G. F. Torres del Castillo}
\email{gtorres@fcfm.buap.mx}
\affiliation{Departamento de F\'isica Matem\'atica, Instituto de Ciencias, Universidad Aut\'onoma de Puebla, 72570 Puebla, Puebla, M\'exico}

\begin{abstract}In this paper we show that the vanishing of the scalar curvature of Ruppeiner-like metrics does not characterize the ideal gas. Furthermore, we claim through an example that flatness is not a sufficient condition to establish the absence of interactions in the underlying microscopic model of a thermodynamic system, which poses a limitation on the usefulness of Ruppeiner's metric and conjecture. Finally, we address the problem of the choice of coordinates in black hole thermodynamics. We propose an alternative energy representation for Kerr-Newman black holes that mimics fully Weinhold's approach. The corresponding Ruppeiner's metrics become degenerate only at absolute zero and have non-vanishing scalar curvatures.
\end{abstract}

\maketitle

\section{Introduction}

Thermodynamics is considered to be one of the finest descriptions of real phenomena. The capability of its systems to predict real-life results in such an accurate fashion has settled it in a privileged place among the accepted physical theories. However, as Gibbs \cite{Gibbs:1928aa} and Carath\'{e}odory \cite{Caratheodory:1909aa} noticed by the end of the 19th century and the beginning of the 20th, respectively, it lacks the mathematical precision of other areas, such as classical mechanics or electrodynamics.

This problem has inspired a line of research whose main goal is to describe thermodynamic systems geometrically. One of the first steps in this direction were the pioneering works of Frank Weinhold \cite{Weinhold:1975aa,Weinhold:1975ab,Weinhold:1975ac,Weinhold:1975ad,Weinhold:1976aa}, where it was shown that Riemannian geometry arises naturally in the study of equilibrium thermodynamics. 

This approach was further explored by George Ruppeiner \cite{Ruppeiner:1979aa}, who proposed a metric (which also arises from thermodynamic principles) conformally equivalent to Weinhold's \cite{Mrugaa:1984aa, Torres-del-Castillo:1993aa}, and suggested a relationship between its scalar curvature and the interactions of the underlying microscopic model, namely, that this scalar is related to the correlation length and that its sign corresponds to whether the inter particle interactions are effectively attractive or repulsive \cite{Ruppeiner:1979aa,Ruppeiner:1995aa,Ruppeiner:2010aa,Ruppeiner:2012aa}. We shall refer to this statement as \emph{Ruppeiner's conjecture}. We set out a particular example where the Ricci scalar of Ruppeiner-like metrics (see definition below) is identically zero in a whole Riemannian submanifold of the space of equilibrium states, except for a set of points where the metric becomes degenerate. This counterexample illustrates that, despite the many tests that the aforementioned conjecture has passed \cite{Ruppeiner:2012aa,Ruppeiner:2012ab,Ruppeiner:2013aa}, its scope of validity has yet to be properly delimited.  

The Riemannian-geometric approach to thermodynamics has been applied to a variety of systems, among which black holes stand out (this approach to black hole thermodynamics was pioneered by Cai et al. \cite{Cai:1999ab}). The purpose of this viewpoint is twofold: on the one hand, it intends to relate the divergences of heat capacities (often regarded as thermodynamic critical points) to geometric singularities of Weinhold's metric structure; on the other, it is aimed to determine the nature of the interactions involved in microscopic models of gravity via Ruppeiner's conjecture. Regarding the former issue, the aforementioned relationship has been established under the usual scheme of black hole thermodynamics, where the mass of black holes is regarded as the analogue of energy, whereas the corresponding \emph{deformation coordinates} are taken to be angular momentum and charge \cite{Mansoori:2013ab,Bravetti:2012ac}. Nonetheless, we share the point of view that this choice is somewhat arbitrary and may be challenged on both mathematical and physical grounds \cite{Medved:2008aa,Shen:2005aa}. Herein, we propose an alternative energy representation for Kerr-Newman black holes that mimics fully Weinhold's geometric approach. With this choice of coordinates, Ruppeiner's metrics become degenerate only at absolute zero and have non-vanishing scalar curvatures (cf. {\AA}man et al.\cite{Aman:2008aa} and Mirza et al. \cite{Mirza:2007aa}).

This paper is organized as follows. In the second section we present a short review on Weinhold's Riemannian-geometric approach. In the third section, we show that the scalar curvature of both Ruppeiner-like and Weinhold's metrics is not a characteristic of thermodynamic systems by exhibiting a family of flat closed hydrostatic systems. Though alluring, we show in the fourth section that the correspondence between microscopic interaction and scalar curvature cannot be fully established, by setting out an example of a flat closed hydrostatic system whose isothermal compressibility goes to infinity on a set of ``critical states''. In section 5 we review the basic concepts of the Riemannian approach in the setting of the thermodynamics of Kerr-Newmann black holes; as said before, the point of view adopted in this section is different from the ones that can be found in the literature. Furthermore, we analyze the behavior of the scalar curvature of Ruppeiner's metric tensor in the Riemannian submanifolds of the space of equilibrium states of this family of black holes. We present some conclusions in section 6.

\section{Hessian Structures in Thermodynamics}

In this section we review the concepts of geometrical thermodynamics introduced by Weinhold \cite{Weinhold:1975aa}.

Let $\Sigma$ be a thermodynamic system and $\E$ the set of all its equilibrium states. Each element of $\E$ is uniquely determined by the values of $n$ deformation coordinates (intrinsic parameters) plus $k$ non-deformation coordinates (extrinsic parameters)\footnote{Though not widely used, the name \emph{deformation coordinates} we adopt here is due to Carath\'eodory \cite{Caratheodory:1909aa}.}, i. e., there exists an injective function

\begin{equation*}
\phi=(\xi^1,\ldots,\xi^k,x^1,\ldots,x^n):\E\to\mathbb{R}^{k+n}.
\end{equation*}

\noindent It is convenient to assume that $\phi(\E)$ is open in $\mathbb{R}^{k+n}$, since $\E$ is thereby endowed with a smooth structure. When $k=1$, the system is called \emph{simple}. We consider only this class of systems.

As a consequence of both the First and Second Laws of Thermodynamics, one can choose either the internal energy or the entropy as the non-deformation parameter that together with $x^1,\ldots,x^n$ constitutes a global chart on $\E$. The former is called the \emph{energy representation} of $\Sigma$, whilst the latter is known as the \emph{entropy representation}.

Another essential postulate for the geometrical approach to thermodynamics is the Entropy Maximum Principle. It asserts that all spontaneous processes (processes for which total work and heat are zero, but the starting and ending points are not the same) reach a final state in which entropy attains a maximum (observe that this processes occur at constant total energy) \cite{Callen:1985aa}. This yields as a result that for any numbers $a^0,\ldots,a^n$, and any $x\in\E$,

\begin{equation*}
a^\alpha a^\beta\left(\frac{\partial^2S}{\partial x^\alpha\partial x^\beta}\right)_x\leq0,
\end{equation*}

\noindent where $\alpha$ and $\beta$ run through $\{0,\ldots,n\}$ and $x^0=U$. Hence, -$\partial_\alpha\partial_\beta S$ are the components of a positive semidefinite tensor, denoted by $g_\text{R}$, which we shall refer to as \emph{Ruppeiner-like metrics} \cite{Ruppeiner:1979aa}. 

It can easily be proven that the Hessian matrix of $U$ with respect to the entropy representation is the matrix representation of a tensor $g_\text{W}$ -- known as \emph{Weinhold's metric tensor} -- which is conformally equivalent to $g_\text{R}$, viz. \cite{Mrugaa:1984aa, Torres-del-Castillo:1993aa}

\begin{equation}
g_\text{W}=Tg_\text{R}\label{eq:eq}.
\end{equation}

\noindent Notice that since $T$ is positive, $g_\text{W}$ is positive semidefinite in $\E$. Hence, any integral distribution transversal to $\ker g_\text{W}$ is the tangent space of a Riemannian submanifold of $\E$. One can easily find such a submanifold in the following way: let $r=\operatorname{rank} g_\text{W}$. Then, by relabeling if necessary,

\begin{equation}
\mathcal{D}=\{x\in\E:x^{r+1}(x)=\text{const.},\ldots,x^n(x)=\text{const.}\}
\end{equation}

\noindent is a Riemannian manifold with the metric $\imath^*g_\text{W}$, where $\imath:\mathcal{D}\to\E$ is the inclusion. In all known cases, every submanifold given by $x^i=\text{const.}$ is a Riemannian submanifold of $\E$, for $i\in\{1,\ldots,n\}$, provided that $\dim\ker g_\text{W}=1$ (observe that $g_\text{R}$ is also a Riemannian metric tensor on these submanifolds).

One appealing feature of this approach to thermodynamics is that it portrays thermodynamic potentials as natural geometric objects: they are potentials of the corresponding dual Hessian structure \cite{Shima:2007aa}. Namely, on $x^i=\text{const.}$, $g_\text{W}=-\tilde\nabla\ed{\tilde U}$, where $\tilde U$ is the thermodynamic potential $y_ix^i$ (no summation) and $\tilde\nabla$ is the flat connection whose affine coordinate system is $(T,y_k)$, with $k\in\{1,\ldots,n\}\setminus\{i\}$, where $y_1,\ldots,y_n$ represent the intensive variables of the system. For instance, Gibbs' Free Energy is the dual potential of internal energy in the submanifold given by constant particle number. This, in turn, makes evident the usefulness of Weinhold's metrics: since thermodynamic systems undergo first-order phase transition whenever the potentials with intensive ``natural variables'' are not concave \cite{Callen:1985aa,Linder:2004aa}, any such system would have states $x\in\mathcal{D}$ such that $\tilde\nabla\ed{\tilde U}_x$ is not negative definite, which amounts to ${g_\text{W}}_x$ not being positive definite. Hence, the behavior of Weinhold's metrics is sensitive to phase transitions, needless of any other auxiliary structure whatsoever (cf. Bravetti et al. \cite{Bravetti:2012ac} and Liu et al.\cite{Liu:2010aa}). 

\section{Flat Hydrostatic Closed Systems}

In this section we consider simple hydrostatic systems with fundamental equation

\begin{equation}
\ed U=T\ed S-p\ed V+\mu\ed N\label{eq:1},
\end{equation}

\noindent where $U$, $T$, $S$, $p$, $V$, $\mu$, and $N$ denote the system's internal energy, temperature, entropy, pressure, volume, chemical potential, and number of particles, respectively. The submanifold $\mathcal{N}$ defined by $N=\text{const.}$ is Riemannian; in terms of coordinates $(T,V)$, the metric tensor $g_\text{R}$ has the form

\begin{equation*}
g_\text{R}=\frac{C_V}{T^2}(\ed T)^2+\frac{1}{\upkappa_TTV}(\ed V)^2.
\end{equation*}

It is well known that a closed ideal gas has a flat manifold of equilibrium states \cite{Nulton:1985aa}. We will prove that this is not a characteristic of an ideal gas: there exist infinitely many flat closed systems, even with $C_V=\text{const.}$ Defining 

\begin{align*}
t &=\log(T),\\v&=2\sqrt{V},
\end{align*}

\noindent the metric $g_\text{R}$ takes the form

\begin{equation}
g_\text{R}=C_V(\ed t)^2+\frac{1}{e^t\upkappa_T}(\ed v)^2.
\end{equation}

\noindent The scalar curvature of $\mathcal{N}$ is zero whenever $\partial (e^t\upkappa_T)^{-1/2}/\partial t$ is a function of $v$ only. Therefore, flat closed systems with constant $C_V$ are given by

\begin{equation}
\upkappa_T^{-1}=e^t\left(tf_1+f_2\right)^2,\label{eq:ic}
\end{equation}

\noindent where $f_1$ and $f_2$ are functions of $v$ only (a similar result holds for Weinhold's metrics, with $\kappa_T^{-1}=(\tilde t\tilde {f_1}+\tilde{f_2})^2$, where $\tilde t:=2\sqrt{C_VT}$, and $\tilde{f_1}$ and $\tilde{f_2}$ are functions of $v$ only). In consequence, the ideal gas is only a particular case of closed hydrostatic system with $C_V=\text{const.}$ and identically vanishing scalar curvature, given by $f_2\propto v^{-1}$ and $f_1=0$ ($\tilde f_1\propto v^{-1}$ and $\tilde f_2=0$, respectively).

We point out that the same holds in the Riemannian submanifold defined by $N=\text{const.}$, which is the case originally analyzed by Ruppeiner \cite{Ruppeiner:1979aa,Ruppeiner:1995aa,Ruppeiner:2010aa,Ruppeiner:2012aa,Ruppeiner:2012ab}. It can readily be verified that Ruppeiner geometry is flat if and only if

\begin{equation}
\left(\frac{\partial\mu}{\partial N}\right)_{T,V}=\frac{T}{N}\left(F_1\log T+F_2\right)^2,\label{eq:ic0}
\end{equation}

\noindent where $F_1$ and $F_2$ are functions of $N$ only. The ideal gas is only the particular case given by $F_1=0$ and $F_2=\text{const.}$

The role that equations \eqref{eq:ic} and \eqref{eq:ic0} play in the ostensible relationship between curvature and microscopic interactions will be revealed below.

\section{Non-interacting systems with first-order phase transitions?}

As we mentioned before, one can construct thermodynamic systems whose space of equilibrium states contains a Riemannian submanifold that is both flat and has critical states. 

For the sake of simplicity, we will consider a simple closed hydrostatic system. We have shown that the isothermal compressibility of one such system with flat space of equilibrium states is given by eq. \eqref{eq:ic}. In particular, let $f_1=c/v^2$ and $f_2=-ct_0/(v_0v)$, where $c$, $v_0$, and $t_0$ are constants. Observe that $\left.g_\text{R}\right|_\mathcal{C}$ is degenerate, where

\begin{equation*}
\mathcal{C}=\left\{x\in\E:\frac{t(x)}{t_0}=\frac{v(x)}{v_0}\right\}.
\end{equation*}

Since $g_\text{W}$ and $g_\text{R}$ are conformally equivalent with a non-vanishing conformal factor (temperature), the degeneracy of $g_\text{R}$ renders $g_\text{W}$ degenerate. Hence, $\mathcal{C}$ is a set of critical states and the system presents first-order phase transitions, yet has non-interacting microscopic constituents (according to Ruppeiner's conjecture).

Notice that a similar statement can be established in the case of Ruppeiner geometry ($N=\text{const.}$, cf. eq. \eqref{eq:ic0}).

This contradictory result brings about the necessity of studying singularities in this context, in such a way that the scope of the correspondence between these and thermodynamic critical points can be properly established. Moreover, it puts a limit on Ruppeiner's conjecture linking the scalar curvature of $g_\text{R}$ to the interactions of the underlying microscopic model.

In the next sections we shall proceed to point out another problem of the Riemannian-geometric approach to thermodynamics, which arises in black hole thermodynamics and is related to the concept of extensive and intensive variables.

\section{Riemannian-geometrical approach to black hole thermodynamics}

Ruppeiner's (and Weinhold's) geometric approach to thermodynamics has been kindled recently in black hole thermodynamics \cite{Aman:2008aa,Banerjee:2010aa,Liu:2010aa,Mansoori:2013ab,Mirza:2007aa,Pidokrajt:2011aa,Ruppeiner:2013aa,Ruppeiner:2008aa,Sahay:2010aa,Shen:2005aa,Tiwari:2008aa}. The main goal of this standpoint is to analyze the scalar curvature of Ruppeiner's metrics for different ``macroscopic'' thermodynamic models of black holes, in order to obtain a hint via Ruppeiner's conjecture of the interactions that a microscopic model of gravity might involve. In this section we analyze Ruppeiner's metrics on the space of equilibrium states of the Kerr-Newmann black hole family $\E$. Appealing to the question of the energy representation in black hole thermodynamics \cite{Medved:2008aa,Shen:2005aa}, we propose an alternative choice of coordinates inspired by the extensive character of entropy in ``ordinary'' thermodynamics, which yields interesting results.

Usually, the triad $(M,L,q)$ -- mass, magnitude of angular momentum, and charge of the black hole, respectively -- is regarded as the black-hole analogue of the energy representation in ``ordinary'' thermodynamics \cite{Belgiorno:2002ab}, with $q\neq 0$ and $L\neq 0$. Despite its wide usage, this common assumption has been challenged \cite{Medved:2008aa}. Moreover, based upon similarity in phase diagrams, it has been suggested that $M$ does not play the role of the internal energy, but is a Legendre transformation of it \cite{Shen:2005aa}. 

The importance of the choice of the energy representation lies in the fact that Ruppeiner's metrics are defined by the Hessian of entropy with respect to the energy representation. Thus, Ruppeiner's metrics depend also on the choice of such a chart as we shall show for the Kerr-Newmann case. We shall use $(x,L,y)$, with $x:=M^2$ and $y:=q^2/2$ (which is not a global chart on the space of equilibrium states of Kerr-Newmann black holes), as the energy representation of the system, since according to Smarr's formula \cite{Davies:1977aa}

\begin{equation}\label{eq:Smarr}
4S=x\left(1-\sqrt{1-\frac{2y}{x}-\frac{L^2}{x^2}}\right)-y,
\end{equation}

\noindent entropy is an extensive function of these variables. Equation \eqref{eq:Smarr} is only valid for black holes satisfying $2y<x$ and $L^2<x^2$\, \cite{Davies:1978aa}. We will further restrict our analysis to real values of $S$. 

The degree-one homogeneity of $S$ implies the existence of a Gibbs-Duhem equation, which means in turn that $g_\text{R}$ is degenerate on $\E$; moreover, any submanifold defined by $x^i=\mathrm{const.}$, for some $i\in\{1,2,3\}$, is Riemannian with metric $\imath^*g_\text{R}$, where $\imath$ is the inclusion. 

To begin, we consider the submanifold defined by $L=\mathrm{const.}\neq0$. Let 

\begin{eqnarray*}
\zeta& =& 2\left(1-\frac{2y}{x}-\frac{L^2}{x^2}\right)^{1/4},\\
\chi&=&\frac{2}{\sqrt{x}}.
\end{eqnarray*}

\noindent In this coordinate chart,

\begin{equation}
{g_\text{R}}_L=\frac{1}{\chi^2}(\ed \zeta)^2+\frac{L^2}{\zeta^2}(\ed\chi)^2,
\end{equation}

\noindent where ${g_\text{R}}_L$ is Ruppeiner's metrics at constant angular momentum. Its scalar curvature is given by
\begin{equation}
R_L=-\frac{4}{L}\left(\frac{\zeta^2}{L\chi^2}+\frac{L\chi^2}{\zeta^2}\right),
\end{equation}

\noindent which diverges only when $\chi$ or $\zeta$ are zero. The former case amounts to a black hole with infinite mass, whereas in the latter, the black hole reaches absolute zero and becomes an extreme black hole (cf. eq. (1.4) in Davies \cite{Davies:1978aa}). This state corresponds to a transition point from an object with horizon into a naked singularity.

We shall now analyze the scalar curvature of Ruppeiner's metrics in the submanifold given by $y=\text{const.}\neq0$. Defining

\begin{equation*}
w=\frac{L}{\sqrt{x(x-2y)}},
\end{equation*}

\noindent one finds that 

\begin{equation}
{g_\text{R}}_q=\frac{\sqrt{x (x-2 y)}}{4 \left(\frac{1}{w^2}-1\right)^{3/2} w^3}(\ed w)^2+\frac{y^2}{4  [x (x-2 y)]^{3/2}w\sqrt{\frac{1}{w^2}-1} }(\ed x)^2.
\end{equation}

The scalar curvature of this metric tensor is

\begin{equation}
R_q=4\frac{ x\left(w^2-1\right)(x-2y)-4 y^2}{Ly^2\sqrt{\frac{1}{w^2}-1} }.
\end{equation}

\noindent Notice that in this case curvature becomes infinite when $\sqrt{w^{-2}-1}=0$, which corresponds to a black hole at absolute zero (cf. $\zeta=0$ in the previous case).

The qualitative change that a black hole suffers as it approaches absolute zero resembles that of thermodynamic systems undergoing phase transitions. Therefore, it is remarkable that the energy representation we have proposed yields a metric whose scalar curvature diverges only at absolute zero on both Riemannian submanifolds of $\E$. Moreover, unlike the common result reported in literature \cite{Banerjee:2010aa,Bravetti:2012ac,Mansoori:2013ab,Shen:2005aa}, it remains finite at points where heat capacities diverge (cf. Ruppeiner \cite{Ruppeiner:2008aa}), whose physical situation is rather unclear \cite{Sokoowski:1980aa}.

\section{Concluding Remarks}

Ever since its appearance in the mid 70's, the Riemannian approach to thermodynamics has remained more a mathematical curiosity than a useful technique to solve actual problems. Ruppeiner's conjecture could be the greatest virtue of this formalism, since it might deliver microscopic information of thermodynamic systems relying solely on macroscopic data. However, as we have illustrated, it may yield wrong information about the underlying microscopic models of thermodynamic systems. We believe that incorporating the study of singularities in the Riemannian-geometric approach to thermodynamics might shed more light on the scope of applicability of this conjecture, turning it into a tool of utmost utility. 

We have also mentioned another unsolved problem of this formalism, which arises in the study of black holes: a criterion for the choice of an appropriate energy representation in an arbitrary thermodynamic system (or equivalently, a coordinate-free definition of Ruppeiner's metric tensor) is still missing. Since the aforementioned choice of coordinates strongly affects the resulting geometric structure, addressing this drawback seems a fundamental task. Besides, its solution might reveal hints to formulate geometric generalizations of thermodynamics. 

\acknowledgments{
The authors thank Bogar D\'iaz for his comments and valuable discussion. This work was supported in part by CONACyT, M\'exico, Grant number 167477-F.
}
%\bibliography{mybib}
%

\end{document}